\documentclass[preprint,superscriptaddress,showpacs]{revtex4-1}

\usepackage{amsmath}
\usepackage{bm}

\begin{document}

\title{Effects of the electron correlation and Breit and hyperfine interactions on the lifetime of the 2p$^5$3s states in neutral neon}

\author{Jiguang Li}
\email{phys\_\,ljg@yahoo.com.cn}
\affiliation{Chimie Quantique et Photophysique, Universit\'e Libre de Bruxelles, B-1050 Brussels, Belgium}
\affiliation{Department of Physics, Lund University, S-221 00 Lund, Sweden}
\author{Per J\"onsson}
\affiliation{School of Technology, Malm\"o University, S-205 06 Malm\"o, Sweden}
\author{Michel Godefroid}
\affiliation{Chimie Quantique et Photophysique, Universit\'e Libre de Bruxelles, B-1050 Brussels, Belgium}
\author{Chenzhong Dong}
\affiliation{College of Physics and Electronic Engineering, Northwest Normal University, 730070 Lanzhou, China}
\author{Gediminas Gaigalas}
\affiliation{Institute of Theoretical Physics and Astronomy, Vilnius University, A. Go\v{s}tauto 12, LT-01108 Vilnius, Lithuania}

\date{\today}

\begin{abstract}
In the framework of the multiconfiguration Dirac-Hartree-Fock method, we investigate the transition properties of four excited states in the $2p^53s$ configuration of neutral neon. The electron correlation effects are taken into account systematically by using the active space approach. The effect of higher-order correlation on fine structures is shown. We also study the influence of the Breit interaction and find that it reduces the oscillator strength of the $^3P^o_1 ~-~ ^1S_0$ transition by 17\%. It turns out that the inclusion of the Breit interaction is essential even for such a light atomic system. Our \textit{ab initio} calculated line strengths, oscillator strengths and transition rates are compared with other theoretical values and experimental measurements. Good agreement is found except for the $^3P^o_2 ~-~ ^1S_0$ M2 transition for which discrepancies of around 15\% between theories and experiments remain. In addition, the impact of hyperfine interactions on the lifetimes of the $^3P^o_0$ and $^3P^o_2$ metastable states is investigated for the $^{21}$Ne isotope (I=3/2). We find that hyperfine interactions reduce the lifetimes drastically. For the $^3P^o_0$ state the lifetime is decreased by a factor of 630.

\end{abstract}

\pacs{31.15.A-, 31.15.ag, 31.30.J-, 32.10.Fn}

\maketitle

\section{Introduction}

Lifetimes of states in the first excited configuration $2p^53s$ for neutral neon are important, not only because of their potential applications~\cite{Matthews1985, Hibbert1993, Hibbert1994, Seaton1998, FroeseFischer2004, Morel2008, Butler2008} relevant to plasma diagnosis, laser techniques, and the interpretation of astrophysical data, but also for investigating electron correlation and relativistic effects or testing many-body theories of atomic physics~\cite{Fielder1979, Fielder1979a, Avgoustoglou1995, Avgoustoglou1998, Savukov2002, Dong2003, Zatsarinny2009}.

A great deal of calculations and measurements have been reported for electric dipole (E1) transition probabilities or corresponding oscillator strengths ($gf$) between the states of the ground $2p^6$ and first excited $2p^53s$ configurations of neutral neon. However, a satisfactory precision has not been achieved yet. For instance, the $gf$ values of the lower $J=1$ level, i.e. $^3P^o_1$ (the LS coupling label is used throughout this paper for convenience), obtained by the nonrelativistic wave functions with relativistic corrections in the Breit-Pauli (BP) approximation range from 0.0102 to 0.0123~\cite{Hibbert1993, Savukov2002, FroeseFischer2004, Zatsarinny2009}, while the relativistic results are larger than 0.0130~\cite{Avgoustoglou1998, Dong2003}. Unfortunately, the inconsistency cannot be resolved by the experimental measurements because of large error bars. The detailed comparisons have been recently reviewed by Chan~\cite{Chan1992}, Avgoustoglou~\cite{Avgoustoglou1998}, Savukov~\cite{Savukov2002} and Zatsarinny~\cite{Zatsarinny2009}.

Another appealing subject is the lifetimes of the two metastable $^3P^o_2$ and $^3P^o_0$ levels in the $2p^53s$ configuration. For isotopes without nuclear spin $I$, the magnetic quadrupole (M2) transition to the ground state is the dominant single-photon decay channel for the $^3P^o_2$ state, while the $^3P^o_0$ level can decay through the magnetic dipole (M1) or electric quadrupole (E2) transition to $^3P^o_{1,2}$ lower states. In 1972, Van Dyck, Johnson, and Shugart measured the composite lifetime of the metastable rare-gas atoms in these two states using the time-of-flight technique~\cite{VanDyck1972}. The experiment sets a lower limit for the lifetime, and the value is 0.8~s for Ne. Recently, Zinner determined the lifetime of the $^3P^o_2$ state by measuring the decay in fluorescence of an ensemble of $^{20}$Ne atoms trapped in a magneto-optical trap (MOT)~\cite{Zinner2003}. It is worth noting, however, that the latest experimental result $\tau = 14.73(14)$~s considerably differs from the earliest theoretical results $\tau=24.4$~s by Small-Warren and Chow Chiu~\cite{Small-Warren1975} and $\tau=29$~s by Fielder, Jr. \textit{et al.}~\cite{Fielder1979, Fielder1979a}. Also it does not agree with recent calculations, that is, 19.8~s by Beck~\cite{Beck2002} with relativistic configuration interaction method, 18.9~s obtained by Desclaux \textit{et al.} (cited in~\cite{Zinner2003}) and Dong \textit{et al.}~\cite{Dong2003} using the multiconfiguration Dirac-Hartree-Fock (MCDHF) method and 16.9~s by Froese Fischer and Tachiev~\cite{FroeseFischer2004} with the multiconfiguration Hartree-Fock (MCHF) method including relativistic corrections in the BP approximation.

On the other hand, for isotopes having a non-zero nuclear spin, issues become complicated since a new decay channel is opened by hyperfine interactions. This transition is referred to as a hyperfine induced transition (HIT) or hyperfine quenching decay mode~\cite{Johnson2011}. Owing to their peculiarity, HITs have attracted much attention during the last several years~\cite{Indelicato1989, Andersson2008a, Li2009, Li2010, Chen2011}, stimulating us to further predict the rates for the $^3P^o_0$ and $^3P^o_2$ metastable states of the $^{21}$Ne isotope.

In this work, we perform large-scale calculations of the transition properties of states in the $2p^53s$ configuration using the GRASP2K package~\cite{Jonsson2007} based on the MCDHF method which allows one to take electron correlation and relativity into account on the same foot. The active space approach is adopted to monitor the convergence of the physical quantities concerned. The importance of the Breit interaction for an accurate determination of the lifetimes is studied. We report the lifetime of these states for abundant isotopes with respect to important decay channels including hyperfine induced transitions.


\section{\label{Sec:TC} Theoretical method and Computational Model}
\subsection{\label{subsec:MCDF} MCDHF method}
The multiconfiguration Dirac-Hartree-Fock method is written up in the monograph by Grant \cite{Grant2007} and we here just give a brief description of the method. Starting from the Dirac-Coulomb Hamiltonian
\begin{eqnarray}
H_{DC} = \sum_{i} \left( c\, {\bm{ \alpha }}_i \cdot {\bm{ p }}_i + (\beta_i -1)c^2 + V^N_i \right) + \sum_{i>j} 1/r_{ij},
\end{eqnarray}
where $V^N$ is the monopole part of the electron-nucleus Coulomb interaction, the atomic state functions (ASFs) describing different
fine structure levels are obtained as linear combinations of
symmetry adapted configuration state functions (CSFs) with same parity $P$, angular momentum $J$ and its $M_J$ component along $z$-direction
\begin{eqnarray}\label{ASF}
\label{ASF} \Psi(P J M_J)  = \sum_{j=1}^{N} c_{j} \Phi(\gamma_{j} P J M_J).
\end{eqnarray}
In Eq. (\ref{ASF}) $c_{j}$ is the mixing coefficient and $\gamma_j$ denotes other appropriate labeling of the configuration state function, for example orbital occupation numbers and coupling tree. The configuration state functions are built from products of one-electron Dirac orbitals. In the self-consistent field (SCF) procedure both the radial parts of the Dirac orbitals and the expansion coefficients are optimized to minimize the energies concerned. Calculations can be performed for a single level, but also for a portion of a spectrum in an extended optimal level (EOL) scheme where optimization is applied on a weighted sum of energies. The Breit interaction
\begin{eqnarray}
\label{eq:Breit}
         B_{ij} =  - \frac{1}{2 r_{ij}}\Biggl[ \bm{\alpha}_{i} \cdot \bm{\alpha}_{j} + \frac{ (\bm{\alpha}_{i} \cdot {\bm{ r_{ij} }})
                (\bm{\alpha}_{j} \cdot {\bm{ r_{ij} }}) } {r_{ij}^2} \Biggr]
\end{eqnarray}
can be further included in subsequent relativistic configuration interaction (RCI) computations.

Once the atomic state functions have been obtained atomic parameters are evaluated in terms of reduced matrix elements of the corresponding tensor operator
\begin{eqnarray}
\langle \,\Psi(P J)\, \| {\bf O}^{(\lambda)} \| \,\Psi(P' J')\, \rangle .
\end{eqnarray}
For the transition, the tensor operator ${\bf O}^{(\lambda)}$ is a multipole radiation field operator. The superscript designates the type of multipole: $\lambda = 1$ for electric multipoles and $\lambda = 0$ for magnetic multipoles. This expectation value reduces to a sum over reduced matrix elements between CSFs by substituting the ASF expansions (\ref{ASF}). Using Racah algebra these reduced matrix elements, in turn, are expressed as a weighted sum over radial integrals involving the radial relativistic one-electron orbitals.

The restriction from Racah algebra that ASFs are built from the same orthogonal radial orbital set can be relaxed by the biorthogonal transformation technique~\cite{Olsen1995, Verdebout2010}. As a result, reduced matrix elements between two atomic state functions described by independently optimized orbital sets can be calculated using standard techniques.

\subsection{\label{subsec:CM} Computational model}
In the framework of the MCDHF method, the building of the configuration space is pivotal not only for capturing the electron correlation effect efficiently, but also for circumventing the convergence problem that one frequently encounters in SCF calculations. In this work, we use the active space (AS) approach to generate the configuration list from the reference configuration set. The reference set is initially made up of (near-)degenerate reference configurations and can be augmented by important CSFs for considering the higher-order correlation effects~\cite{Jonsson2010, Carette2011a, Carette2011}. We name in this paper the initial set as MR$^{(0)}$ and the latter MR$^{(1)}$. More generally, the reference set MR can be divided into several subsets for explaining correlation effects between specific electron pairs. According to the perturbation theory, the first-order correction of ASFs is expressed as a linear combination of CSFs that are obtained by replacing one or two occupied orbitals of the reference configurations in MR$^{(0)}$ with active orbitals~\cite{FroeseFischer1997}. The set of active orbitals is enlarged systematically, which makes it possible to monitor the convergence of the physical quantities under investigation. Higher-order correlation corrections are more difficult to deal with since the number of CSFs grows rapidly and easily goes beyond the capability of even a large computer system. Yet most of CSFs actually make fractional contributions to ASFs. The key point in this step is to define the MR$^{(1)}$ appropriately. In general, significant CSFs in first-order correction are added to the MR$^{(0)}$ to form the MR$^{(1)}$ set. The configuration space is further expanded by single(S) and double(D) replacements for orbitals of CSFs belonging to MR$^{(1)}$ with the ones appearing in a given active set.

For the case of neon, Lee \textit{et al.} pointed out that higher-order correlations of the L shell are significant for the ground state~\cite{Lee1971}. Afterwards, Dong \textit{et al.} also showed by MCDHF calculations that CSFs generated from the $2s^22p^43p^2$ configuration improves the accuracy of the transition rates to a great extent~\cite{Dong2003}. As a result, we choose $\{\{2s^22p^6; 2s^22p^43p^2; 2s^22p^53p\}\,; \{1s^22s^22p^6\}\}$ as the MR$^{(0)}$ set for the ground state and $\{\{2s^22p^53s\}\,;\{1s^22s^22p^53s\}\}$ for the four lowest excited states, respectively. The first subsets in MR$^{(0)}$ aim at accounting for the outer electron correlations and the second for correlations involving the $1s$ core. The construction of the configuration space is presented in table \ref{NCSFs}. As can be seen from this table, these correlation models are marked with $nl$SD where $n$ and $l$, if appearing, designate, respectively, the maximum principal and orbital angular momentum quantum numbers of the active orbitals. The core correlation involving $1s$ electrons (labeled as ``CC'') is taken into account by allowing SD excitations from the $1s$ core to the largest active set. To incorporate the residual higher-order correlations of outer shells, the $\{2s2p^53s3p\,; 2s^22p^43d^2\,; 2s2p^63s\}$ configurations are added to the first subset of MR$^{(0)}$ for the ground state and $\{2s^22p^33s3p^2\,; 2s2p^53s3d\,; 2s^22p^33s3d^2\,; 2s^22p^43s3p\,; 2s^22p^53d\}$ for the excited states to set up MR$^{(1)}$. The SD excitations up to $n=4$ are based on the MR$^{(1)}$ and the CSFs are appended to the CC model to form the final configuration spaces (marked with MR). It is worth noting that the addition of $2s^22p^53p\,; 2s2p^63s\,; 2s^22p^43s3p\,; 2s^22p^53d$ configurations in the reference sets is ascribed to the requirement of closing the CSF space under de-excitation by the biorthogonal transformation technique~\cite{Jonsson2007}.

In practice, we further eliminate the CSFs that do not interact with reference configurations~\cite{Jonsson2007, FroeseFischer1997} in order to raise the calculation efficiency. As can be seen from table \ref{NCSFs}, the number of CSFs of the reduced configuration space is considerably smaller than the corresponding full one. These removed CSFs contribute to the atomic properties under investigation at higher-order and the quantitative influence can be seen in table \ref{HCE}. Using 9fSD and CC models, we calculate excitation energies and $^3P^o_1 ~-~ ^1S_0$ line strength. It is found that the impact of removed CSFs on excitation energies between different configurations or terms is fractional whereas remarkably large for the fine structure splitting. For example, the influence reaches around 20\% for the $^3P^o_1 ~-~ ^3P^o_0$ fine structure. Comparing the E1 line strengths obtained with the two configuration spaces, we see that the loss of CSFs contributes to the weak line by 3\% but merely 0.2\% for the strong line.

\begin{table}[!ht]
\begin{scriptsize}
\caption{\label{NCSFs} The number of configuration wave functions (CSFs) in various correlation models. $J^P$ are the total angular momentum ($J$) and parity ($P$) of an atomic state. MR stands for the reference configuration set, AO for the set of active orbitals. The number of CSFs without reduction is presented in the parentheses following the number of the reduced configuration space.  * indicates that all active orbitals are included.}
\begin{tabular}{cccccccccc}
\hline
\hline
                                       &              &        & \multicolumn{3}{c}{NCSF}       \\
\cline{4-6}         MR                 &    AO        &  Model & \multicolumn{3}{c}{$J^P=0^e$}  \\
\hline
$\{2s^22p^6\,; 2s^22p^43p^2\,; 2s^22p^53p\}$ &                              & DF     &  \multicolumn{3}{c}{12(12)}         \\
                                       & \{3s,3p,3d\}                       & 3SD    &  \multicolumn{3}{c}{549(728)}       \\
                                       & \{3*,4s,4p,4d,4f\}                 & 4SD    &  \multicolumn{3}{c}{3731(6021)}     \\
                                       & \{3*,4*,5s,5p,5d,5f,5g\}           & 5SD    &  \multicolumn{3}{c}{10884(19355)}   \\
                                       & \{3*,4*,5*,6s,6p,6d,6f,6g,6h\}     & 6SD    &  \multicolumn{3}{c}{23166(43967)}   \\
                                       & \{3* 4*,5*,6*,7s,7p,7f,7f\}        & 7fSD   &  \multicolumn{3}{c}{35746(67433)}   \\
                                       & \{3*,4*,5*,6*,7*,8s,8p,8d,8f\}     & 8fSD   &  \multicolumn{3}{c}{51122(96017)}   \\
                                       & \{3*,4*,5*,6*,7*,8*,9s,9p,9d,9f\}  & 9fSD   &  \multicolumn{3}{c}{69294(129719)}  \\
$\bigcup \{1s^22s^22p^6\}$             & \{3s,\ldots,9f\}                   & CC     &  \multicolumn{3}{c}{71406(132005)}  \\
$\bigcup\{2s2p^53s3p\,; 2s^22p^43d^2\,;$  & \{3s,3p,3d,4s,4p,4d,4f\}        & MR &  \multicolumn{3}{c}{81327(143037)}  \\
$2s^22p^53p\,; 2s2p^63s\}$                                                                   \\[0.2cm]
\hline
                                       &                          &        &  $J^P=0^o$ & $J^P=1^o$ & $J^P=2^o$ \\
\hline
$\{2s^22p^53s\}$                       &                                   & DF     &   1(1)         &  2(2)          & 1(1)           \\
                                       & \{3s,3p,3d\}                      & 3SD    &   86(145)      &  326(369)      & 287(431)       \\
                                       & \{3*,4s,4p,4d,4f\}                & 4SD    &   444(866)     & 1942(2279)     & 1821(2887)     \\
                                       & \{3*,4*,5s,5p,5d,5f,5g\}          & 5SD    &   1192(2495)   & 5500(6734)     & 5327(9027)     \\
                                       & \{3*,4*,5*,6s,6p,6d,6f,6g,6h\}    & 6SD    &   2442(5325)   & 11600(14639)   & 11469(20435)   \\
                                       & \{3* 4*,5*,6*,7s,7p,7f,7f\}       & 7fSD   &   3727(8162)   & 17846(22332)   & 17611(30868)   \\
                                       & \{3*,4*,5*,6*,7*,8s,8p,8d,8f\}    & 8fSD   &   5289(11619)  & 25468(31683)   & 25100(43485)   \\
                                       & \{3*,4*,5*,6*,7*,8*,9s,9p,9d,9f\} & 9fSD   &   7128(15696)  & 34466(42692)   & 33936(58286)   \\
$\bigcup \{1s^22s^22p^53s\}$           & \{3s,\ldots,9f\}                  & CC     &   11744(30740) & 59320(83520)   & 55901(113950)  \\
$\bigcup \{2s^22p^33s3p^2\,; 2s2p^53s3d\,;$ & \{3s,3p,3d,4s,4p,4d,4f\}     & MR     &   45368(63831) & 135830(173967) & 187309(238761) \\
$2s^22p^33s3d^2\,; 2s^22p^43s3p\,; 2s^22p^53d\}$                                  \\[0.2cm]
\hline
\hline
\end{tabular}
\end{scriptsize}
\end{table}

\begin{table}[!ht]
\begin{footnotesize}
\caption{\label{HCE} Comparison of transition energies (in cm$^{-1}$) and E1 line strengths (in a.u.) obtained with reduced (r) and full (f) 9fSD and CC configuration spaces. B: Babushkin gauge; C: Coulomb gauge. NIST data~\cite{NIST} are presented for reference.}
\begin{tabular}{cccccccccccccccc}
\hline
\hline
        &  \multicolumn{4}{c}{Excitation energy} && \multicolumn{2}{c}{Line strength ($^3P^o_1 ~-~ ^1S_0$)} && \multicolumn{2}{c}{Line strength ($^1P^o_1 ~-~ ^1S_0$)}    \\
\cline{2-5}\cline{7-8}\cline{10-11} Model & $^1S_0 ~-~ ^3P^o_1$ & $^3P^o_1 ~-~ ^1P^o_1$ & $^3P^o_1 ~-~ ^3P^o_0$ & $^3P^o_2 ~-~ ^3P^o_0$ &&  B   &  C   && B  &  C \\
\hline
9fSD(r)   & 134838 &  1391.95   &   475.16   &   874.87  && 0.03417  &  0.03419  &&  0.3562 &  0.3556 \\
9fSD(f)   & 134837 &  1397.68   &   400.33   &   835.33  && 0.03512  &  0.03508  &&  0.3555 &  0.3550 \\
\hline
CC (r)    & 135398 &  1381.44   &   479.10   &   880.12  &&  0.03488 &  0.03410  &&  0.3524 &  0.3440 \\
CC (f)    & 135395 &  1387.28   &   404.84   &   840.76  &&  0.03583 &  0.03499  &&  0.3517 &  0.3435 \\[0.2cm]
NIST      & 134459 & 1429.43    &   359.35   &   776.80  &&          &           &&         &         \\
\hline
\hline
\end{tabular}
\end{footnotesize}
\end{table}

\subsection{\label{subsec:BI} Breit Interaction}
Ynnerman \textit{et al.} have demonstrated that the Breit interaction plays a key role in the spin-forbidden $^3P^o_1 ~-~ ^1S_0$ transition of low-Z Be-like ions~\cite{Ynnerman1995}. Avgoustoglou \textit{et al.} have also illustrated the effect of the Breit interaction on the Ne I transition energies~\cite{Avgoustoglou1995}. In this subsection, we investigate the Breit interaction effect on transition energies and on the $^3P^o_1 ~-~ ^1S_0$ line strength. It should be emphasized that the full configuration space must be used because the strategy adopted for reducing the number of CSFs does not apply to the Dirac-Breit Hamiltonian. As examples, we present results with and without the Breit interaction, which are obtained using the full DF, 9fSD and CC configuration models. As can be seen from this table, the Breit interaction substantially affects the physical quantities concerned. For instance, the impact of the Breit interaction on the line strength for the $^3P^o_1 ~-~ ^1S_0$ transition reaches about 17\%.

\begin{table}[!ht]
\caption{\label{BE} Breit interaction effect on the transition energies and the $^3P^o_1 ~-~ ^1S_0$ line strength. The full configuration space is used in these calculations. B: Babushkin gauge; C: Coulomb gauge. NIST data~\cite{NIST} are presented for references.}
\begin{tabular}{lccccccccccccccc}
\hline
\hline
        &  \multicolumn{3}{c}{Fine structures (in cm$^{-1}$)} && \multicolumn{2}{c}{Line strength ($^3P^o_1 ~-~ ^1S_0$) (in a.u.)}      \\
\cline{2-4}\cline{6-7} Model & $^3P^o_1 ~-~ ^1P^o_1$ & $^3P^o_1 ~-~ ^3P^o_0$ & $^3P^o_2 ~-~ ^3P^o_0$ &&  B   &  C    \\
\hline
DF                  &  1400.06   &   389.16   &   820.83  && 0.02584  &  0.03172  \\
DF + Breit          &  1485.00   &   347.73   &   752.63  && 0.02184  &  0.02684  \\
\hline
9fSD                &  1397.68   &   400.33   &   835.33  && 0.03512  &  0.03508  \\
9fSD + Breit        &  1381.77   &   358.59   &   767.36  && 0.02989  &  0.02986  \\
\hline
CC                  &  1387.28   &   404.84   &   840.76  &&  0.03583 &  0.03499  \\
CC + Breit          &  1370.59   &   362.71   &   772.65  &&  0.03054 &  0.02979  \\[0.2cm]
NIST                &  1429.43   &   359.35   &   776.80  &&          &           \\
\hline
\hline
\end{tabular}
\end{table}

\section{\label{RD} Results and Discussion}

\subsection{\label{subsec:EE} Excitation Energies}
As functions of the computational models described in Section \ref{subsec:CM}, the excitation energies are presented in table \ref{EE}. The reduced configuration space is used at each step except for the last one where the Breit interaction is accounted for with the full configuration list. It is found that the correlation between outer electrons is saturated in the 9fSD model. The core correlation and high-order effect make relatively small contributions but significant to bring the excitation energies to a satisfactory agreement with the experimental values~\cite{NIST}. Compared with other theories, the present excitation energies between ground and excited states are better than those obtained by Avgoustoglou \textit{et al.}~\cite{Avgoustoglou1995,Avgoustoglou1998} and by Savukov \textit{et al.}~\cite{Savukov2002} with many-body perturbation theory (MBPT), but are not as excellent as the MCDHF data of Dong \textit{et al.}~\cite{Dong2003} and MCHF values of Froese Fischer and Tachiev~\cite{FroeseFischer2004}. It should be pointed out, however, that the core excitations have been neglected in these two calculations. Moreover, in the work of Froese Fischer and Tachiev relativistic effects were included through the Breit-Pauli Hamiltonian but the orbit-orbit interaction, which is part of the Breit interaction, is ignored. In addition, we noticed that present calculated fine structure splittings are consistent with the experimental values~\cite{NIST}, and are better than other calculations as well.

\begin{table}
\caption{\label{EE} Excitation energies (in cm$^{-1}$) of $2p^53s$ levels for neutral neon.}
\begin{tabular}{ccccccccccccccccc}
\hline
\hline
Model & $^3P^o_2$ & $^3P^o_1$ & $^3P^o_0$ & $^1P^o_1$ & $^3P^o_1 ~-~ ^1P^o_1$ & $^3P^o_1 ~-~ ^3P^o_0$ & $^3P^o_2 ~-~ ^3P^o_0$ \\
\hline
DF     & 140733  & 141165  & 141554  & 142565  & 1400 & 389 & 821 \\
3SD    & 138180  & 138599  & 138986  & 139977  & 1378 & 387 & 806 \\
4SD    & 133797  & 134203  & 134652  & 135600  & 1397 & 449 & 854 \\
5SD    & 134420  & 134821  & 135288  & 136207  & 1386 & 468 & 869 \\
6SD    & 134572  & 134972  & 135446  & 136352  & 1380 & 474 & 875 \\
7fSD   & 134494  & 134894  & 135370  & 136277  & 1383 & 475 & 876 \\
8fSD   & 134462  & 134861  & 135336  & 136253  & 1392 & 475 & 874 \\
9fSD   & 134438  & 134838  & 135313  & 136230  & 1392 & 475 & 875 \\
CC     & 134997  & 135398  & 135877  & 136780  & 1382 & 479 & 880 \\
MR     & 134347  & 134783  & 135191  & 136173  & 1390 & 408 & 845 \\
Breit  & 134356  & 134765  & 135127  & 136141  & 1375 & 362 & 771 \\[0.2cm]
\multicolumn{8}{c}{\textbf{Others}}                \\
Avgoustoglou \textit{et al.} \cite{Avgoustoglou1995}  & 134011  & 134406 & 134757 & 135570 & 1164 & 351 & 746 \\
Avgoustoglou \textit{et al.} \cite{Avgoustoglou1998}  & & 133770 & & 135196 \\
Savukov \textit{et al.}  \cite{Savukov2002} & & 132738 && 134231 \\
Dong \textit{et al.} \cite{Dong2003} & 134110 & 134567 & 134940 & 135969 & 1402 & 373 & 830 \\
Froese Fischer and Tachiev \cite{FroeseFischer2004} & 134038 & 134452 & 134807 & 135887 & 1435 & 355 & 769 \\
NIST \cite{NIST} & 134042  & 134459  & 134819  & 135889  & 1430 & 360 &  777   \\[0.2cm]
\hline
\hline
\end{tabular}
\end{table}

\subsection{The $^{3,1}P^o_1 ~-~ ^1S_0$ E1 transitions}
In table \ref{E1 tran} we report line strengths ($S$) and corresponding oscillator strengths ($gf$) in Babushkin and Coulomb gauges for $^{3,1}P^o_1 ~-~ ^1S_0$ transitions. These two gauges are related to the nonrelativistic length and velocity form of transition operators, respectively~\cite{Grant1974}. The convergence of line strengths and oscillator strengths and the good consistency found between the two gauges further justify our computational models and suggest reliable atomic wave functions.

Theoretical and experiments values published during the last two decades are also displayed in table \ref{E1 tran}. For the $^3P^o_1 ~-~ ^1S_0$ transition, we see an excellent agreement with the semi-empirical calculations of Hibbert \textit{et al.}~\cite{Hibbert1993} and of Seaton~\cite{Seaton1998}. The present $gf$ value differs from MBPT values of Avgoustoglou \textit{et al.}~\cite{Avgoustoglou1998} and of Savukov \textit{et al.}~\cite{Savukov2002} by 30\% and 18\%, respectively. Such large discrepancies might be attributed to the Breit interaction that was completely or partly neglected in MBPT calculations of transition properties. Good agreement is found with the results obtained by Dong \textit{et al.}~\cite{Dong2003}. They adopted L\"owdin's approach~\cite{Lowdin1955} to account for non-orthogonal orbitals in transitions~\cite{Fritzsche2000, Fritzsche2012} instead of the biorthogonal transformation technique used in this work. The difference between the results of Froese Fischer and Tachiev~\cite{FroeseFischer2004} and ours is about 12\%. Using the similar Breit-Pauli approximation to Froese Fischer and Tachiev, Zatsarinny and Bartschat recently calculated the $gf$ values by the B-spline method~\cite{Zatsarinny2009}, whose results approach our calculations. Compared with experimental measurements, our results perfectly agree with Zhong \textit{et al.} and are in good consistency with Chan \textit{et al.}~\cite{Chan1992} and Suzuki \textit{et al.}~\cite{Suzuki1994} with respect to the experimental errors. It is worth noting that all these experiment measurements in good agreement with present calculations were obtained by the electron-energy-loss spectrometer (EELS) method.

For the $^1P^o_1 ~-~ ^1S_0$ transition, the agreement between theories and experiments is better than for the spin-forbidden transition. But we find that the semi-empirical results of Hibbert \textit{et al.}~\cite{Hibbert1993} and of Seaton~\cite{Seaton1998} and the B-spline values by Zatsarinny and Bartschat~\cite{Zatsarinny2009} are larger than other theoretical data. Present $gf$ is also consistent with all experimental results listed in this table except for the value of Curtis \textit{et al.}~\cite{Curtis1995}.
\begin{table}
\begin{scriptsize}
\centering
\caption{\label{E1 tran} The convergence trends of line strengths $S$ (in a.u.) and corresponding oscillator strengths $gf$ for the $^{1,3}P^o_{1} ~-~ ^1S_0$ E1 transitions of neutral neon. B: Babushkin gauge, C: Coulomb gauge.}
\begin{tabular}{cccccccccccccccccccccccc}
\hline
\hline
&& \multicolumn{5}{c}{$^3P^o_1 - ^1S_0$} && \multicolumn{5}{c}{$^1P^o_1 - ^1S_0$}  \\
\cline{3-7}\cline{9-13} && \multicolumn{2}{c}{$S$} && \multicolumn{2}{c}{$gf$}  && \multicolumn{2}{c}{$S$} && \multicolumn{2}{c}{$gf$}  \\
\cline{3-4}\cline{6-7}\cline{9-10}\cline{12-13} Model &&  B & C && B & C  && B & C && B & C  \\
\hline
DF     && 0.03172 & 0.02584 && 0.01360 & 0.01108 && 0.3428 & 0.2796 && 0.1484 & 0.1211 \\
3SD    && 0.03199 & 0.03162 && 0.01347 & 0.01331 && 0.3528 & 0.3484 && 0.1500 & 0.1482 \\
4SD    && 0.03250 & 0.03315 && 0.01325 & 0.01351 && 0.3474 & 0.3539 && 0.1431 & 0.1458 \\
5SD    && 0.03327 & 0.03352 && 0.01363 & 0.01373 && 0.3469 & 0.3487 && 0.1435 & 0.1443 \\
6SD    && 0.03383 & 0.03378 && 0.01387 & 0.01385 && 0.3485 & 0.3469 && 0.1443 & 0.1437 \\
7fSD   && 0.03393 & 0.03394 && 0.01390 & 0.01391 && 0.3509 & 0.3501 && 0.1453 & 0.1449 \\
8fSD   && 0.03415 & 0.03419 && 0.01399 & 0.01401 && 0.3552 & 0.3546 && 0.1470 & 0.1468 \\
9fSD   && 0.03417 & 0.03419 && 0.01400 & 0.01400 && 0.3562 & 0.3556 && 0.1474 & 0.1472 \\
CC     && 0.03488 & 0.03410 && 0.01435 & 0.01402 && 0.3524 & 0.3440 && 0.1464 & 0.1429 \\
MR     && 0.03579 & 0.03557 && 0.01465 & 0.01456 && 0.3527 & 0.3504 && 0.1459 & 0.1449 \\
Breit  && 0.03032 & 0.03007 && 0.01241 & 0.01231 && 0.3583 & 0.3556 && 0.1482 & 0.1471 \\
\hline
\multicolumn{13}{c}{\textbf{Theories}}\\
Hibbert \textit{et al.} \cite{Hibbert1993}               &&  &  &&    0.0123  &                    && & &&   0.1607  &                    \\
Seaton \cite{Seaton1998}                                 &&  &  &&  \multicolumn{2}{c}{0.0126}     && & && \multicolumn{2}{c}{0.168}      \\
Avgoustoglou \textit{et al.} \cite{Avgoustoglou1998}     &&  &  &&    0.0163  & 0.0156             && & &&   0.161   &   0.147            \\
Savukov \textit{et al.}  \cite{Savukov2002}              &&  &  &&  \multicolumn{2}{c}{0.0102}     && & && \multicolumn{2}{c}{0.1459}     \\
Dong \textit{et al} \cite{Dong2003}                      && 0.03175 & 0.03309 && 0.01298  & 0.01353 && 0.3492 & 0.3587 && 0.1442 & 0.1482 \\
Froese Fischer and Tachiev \cite{FroeseFischer2004}      && 0.02680 &  &&    0.01095 &             && 0.3668 & &&   0.1514  &                \\
Zatsarinny and Bartschat \cite{Zatsarinny2009}           &&  &  &&    0.0118  & 0.0116             && & &&   0.159   &   0.156            \\
\hline
\multicolumn{13}{c}{\textbf{Experiments}}\\
Chan \textit{et al.} \cite{Chan1992}                     &&  &  &&  \multicolumn{2}{c}{0.0118(6)}   && & && \multicolumn{2}{c}{0.159(8)}   \\
Ligtenberg \textit{et al.} \cite{Ligtenberg1994}         &&  &  &&  \multicolumn{2}{c}{0.01017(30)} && & && \multicolumn{2}{c}{0.1369(35)} \\
Suzuki \textit{et al.} \cite{Suzuki1994}                 &&  &  &&  \multicolumn{2}{c}{0.0106(14)}  && & && \multicolumn{2}{c}{0.137(18)}  \\
Curtis \textit{et al.} \cite{Curtis1995}                 &&  &  &&  \multicolumn{2}{c}{0.0084(3)}   && & && \multicolumn{2}{c}{0.165(11)}  \\
Gibason \textit{et al.} \cite{Gibson1995}                &&  &  &&  \multicolumn{2}{c}{0.01095(32)} && & && \multicolumn{2}{c}{0.1432(38)} \\
Zhong \textit{et al.} \cite{Zhong1997}                   &&  &  &&  \multicolumn{2}{c}{0.0124(38)}  && & && \multicolumn{2}{c}{0.156(9)}   \\
\hline
\hline
\end{tabular}
\end{scriptsize}
\end{table}

\subsection{\label{subsec:M2 tran} The $^3P^o_2 ~-~ ^1S_0$ M2 transition}
In table \ref{M2 tran} we display the $^3P^o_2 ~-~ ^1S_0$ M2 transition rates and corresponding line strengths as functions of the computational models as well as other theoretical and experimental values when available. It is found that our results are in good consistency with results of Beck \cite{Beck2002}, Dong \textit{et al.}~\cite{Dong2003}, Desclaux \textit{et al.}~\footnote{This value is cited in Ref.~\cite{Zinner2003}} and Froese Fischer and Tachiev~\cite{FroeseFischer2004}. However, all theoretical predictions differ from the experimental value~\cite{Zinner2003} by amounts ranging from 14\% to 40\%. To explain such large discrepancies, further experiments are called for.

\begin{table}[!ht]
\caption{\label{M2 tran} Line strengths $S$ (in a.u.) and rates $A$ (in s$^{-1}$) for the $^3P^o_2 ~-~ ^1S_0$ M2 transition as a function of the active space. Numbers in square brackets stand for the power of 10, and in parentheses for error bars. }
\begin{tabular}{cccccc}
\hline
\hline
Model  &  $S$  &  $A$ \\
\hline
DF     & 3.766 &  6.199[-2] \\
3SD    & 3.730 &  5.602[-2] \\
4SD    & 3.916 &  5.006[-2] \\
5SD    & 4.031 &  5.275[-2] \\
6SD    & 4.159 &  5.473[-2] \\
7fSD   & 4.228 &  5.548[-2] \\
8fSD   & 4.332 &  5.678[-2] \\
9fSD   & 4.350 &  5.697[-2] \\
CC     & 4.284 &  5.727[-2] \\
MR     & 4.335 &  5.657[-2] \\
Breit  & 4.345 &  5.672[-2] \\[0.2cm]
\multicolumn{3}{c}{Theories} \\
Small-Warren and Chow Chiu~\cite{Small-Warren1975} & & 4.10[-2]   \\
Indelicato \textit{et al.}~$^{\dagger}$ &  &  4.55[-2] \\
Beck \cite{Beck2002} &  & 5.05[-2] \\
Dong \textit{et al.}~\cite{Dong2003}&  & 5.29[-2] \\
Desclaux \textit{et al.}~$^{\ddag}$ &   &  5.29[-2] \\
Froese Fischer and Tachiev~\cite{FroeseFischer2004} & 4.525 & 5.838[-2] \\
\multicolumn{3}{c}{Experiments} \\
Zinner \textit{et al.}~\cite{Zinner2003} & & 0.06790(64) \\
\hline
\hline
\end{tabular}\\
$^{\dagger}$ This values is cited in Ref.~\cite{Walhout1994};\\
$^{\ddag}$ This values is cited in Ref.~\cite{Zinner2003}.
\end{table}

\subsection{\label{subsec:M1E2} The $^3P^o_0 ~-~ ^3P^o_1$ M1 and $^3P^o_0 ~-~ ^3P^o_2$ E2 transitions}
Line strengths and rates for $^3P^o_0 ~-~ ^3P^o_1$ M1 and $^3P^o_0 ~-~ ^3P^o_2$ E2 transitions are presented in table \ref{M1E2 tran} with the corresponding transition energies. For the M1 transition, we note that the rate is much more sensitive to the transition energy than to the line strength that hardly changes with the computational models. As a result, higher-order electron correlation and the Breit interaction must be taken into account to achieve high accuracy for the M1 transition rate due to their considerably effects on fine structures as discussed in Sec. \ref{subsec:CM} and Sec. \ref{subsec:BI}. It is found from table \ref{M1E2 tran} that our final result is in good agreement with other theoretical calculations.

For the E2 transition the rate is five orders of magnitude smaller than the M1 transition, and thus is negligible. However, we discovered that the transition probabilities in Babushkin and Coulomb gauges are not consistent with each other even with large configuration spaces. As can be seen from table \ref{M1E2 tran}, the inconsistency arises from the deviation of line strengths in Coulomb gauge from those in Babushkin gauge, although they converge with the expansion of configuration space. A strongly gauge-dependency of transition probabilities has also been found in the preceding investigation on the spin-forbidden $2s2p~^3P^o_1 ~-~ 2s^2~^1S_0$ transition of the Be-like C ion~\cite{Ynnerman1995, Jonsson1998}. Chen \textit{et al.} explained that this gauge dependency is caused by the neglect of the negative-energy state which significantly influence the velocity-gauge results~\cite{Chen2001}. Therefore, we argue that the gauge dependency of the E2 transition rate in the case of Ne is brought about for the same reason.

\begin{table}
\caption{\label{M1E2 tran} Line strengths $S$ (in a.u.) and rates $A$ (in s$^{-1}$) together with corresponding transition energies (in cm$^{-1}$) of the $^3P^o_0 ~-~ ^3P^o_1$ M1 and $^3P^o_0 ~-~ ^3P^o_2$ E2 transitions for neon. $\triangle$E represents transition energy. B: Babushkin gauge; C: Coulomb gauge. The number in square brackets represents the power of 10.}
\begin{tabular}{cccccccccccccc}
\hline
\hline
& \multicolumn{3}{c}{M1} && \multicolumn{5}{c}{E2} \\
\cline{2-4}\cline{6-10} Model & $\triangle E$ & $S$ & $A$ && $\triangle E$ & $S_B$ & $S_C$ & $A_B$ & $A_C$ \\
\hline
DF     & 389 & 1.835 & 2.917[-3] && 821 & 3.91[-1] & 1.26     & 1.63[-8] & 5.27[-8]  \\
3SD    & 387 & 1.838 & 2.871[-3] && 806 & 4.21[-1] & 1.61[-1] & 1.60[-8] & 6.14[-9]  \\
4SD    & 449 & 1.833 & 4.462[-3] && 854 & 4.10[-1] & 6.61[-2] & 2.09[-8] & 3.37[-9]  \\
5SD    & 468 & 1.829 & 5.049[-3] && 869 & 4.12[-1] & 1.28[-1] & 2.28[-8] & 7.12[-9]  \\
6SD    & 474 & 1.828 & 5.258[-3] && 875 & 3.90[-1] & 1.65[-1] & 2.24[-8] & 9.48[-9]  \\
7fSD   & 475 & 1.828 & 5.293[-3] && 876 & 3.72[-1] & 1.67[-1] & 2.15[-8] & 9.61[-9]  \\
8fSD   & 475 & 1.830 & 5.279[-3] && 874 & 3.20[-1] & 2.94[-2] & 1.82[-8] & 1.68[-9]  \\
9fSD   & 475 & 1.830 & 5.296[-3] && 875 & 3.09[-1] & 1.14[-2] & 1.77[-8] & 6.53[-10] \\
CC     & 479 & 1.825 & 5.414[-3] && 880 & 3.14[-1] & 3.83[-4] & 1.86[-8] & 2.27[-11] \\
MR     & 408 & 1.821 & 3.348[-3] && 845 & 3.15[-1] & 1.47[-3] & 1.51[-8] & 7.08[-11] \\
Breit  & 362 & 1.849 & 2.358[-3] && 771 & 3.14[-1] & 1.90[-3] & 9.61[-9] & 5.81[-11] \\[0.2cm]
NIST   & 359 &       &           && 777 &          &          &          &           \\
\multicolumn{10}{c}{Theory} \\
Small-Warren \textit{et al.} \cite{Small-Warren1975} &  & & 2.33[-3] && & & & & \\
Dong \textit{et al.} \cite{Dong2003}   & & & 2.308[-3] && & & & & \\
Froese Fischer and Tachiev~\cite{FroeseFischer2004} & 355 & 1.864 & 2.240[-3] && & & & & \\

\hline
\hline
\end{tabular}
\end{table}

\subsection{\label{subsec:HIT} Hyperfine induced $^3P^o_{0,2} ~-~ ^1S_0$ E1 transitions}
In the presence of hyperfine interactions, the electronic angular momentum $J$ is coupled with the nuclear angular momentum $I$ to form the total angular momentum $F$ of the atomic system and only the latter is the good quantum number. As a result, new decay channels can be opened by hyperfine interactions, which affect lifetimes of metastable states substantially. These transitions, named as hyperfine induced transitions, have been investigated extensively during the last decade owing to their potential applications in many fields~\cite{Indelicato1989, Andersson2008a, Li2009, Li2010, Chen2011, Johnson2011}. Neon possesses a stable isotope $^{21}$Ne with nuclear spin $I=3/2$, a magnetic dipole moment $\mu_I$=$-$0.661797 n.m. and with an electric quadrupole moment Q=0.103 barns in the nuclear ground state~\cite{Stone2005}. Two E1 transitions from the metastable states $^3P^o_{0,2}$ to the ground state $^1S_0$ can be induced by hyperfine interactions in $^{21}$Ne isotope. In this section, we predict the decay rates of these two transitions.

Methods calculating the HIT rate have been reviewed in Ref. \cite{Johnson2011}. Based on perturbation theory, the HIT rate of $^{21}$Ne can be estimated by
\begin{equation}
A = \frac{2.02613 \times 10^{18}}{3 \lambda^3} S_{HIT},
\end{equation}
where $\lambda$ is the HIT transition wavelength in ${\rm \AA}$, $S_{HIT}$ the corresponding line strength that is expressed as
\begin{equation}\label{SHIT}
S_{HIT} = |h_1 \langle ^3P^o_1 ||{\bf O}^{(1)}|| ^1S_0 \rangle + h_2 \langle ^1P^o_1 ||{\bf O}^{(1)}|| ^1S_0 \rangle |^2.
\end{equation}
For the latter equation, we only take into account the effect of the adjacent $^3P^o_1$ and $^1P^o_1$ perturbative states. The two reduced matrix elements appearing in Eq. (\ref{SHIT}) are the square roots of line strength $S$ presented in table \ref{E1 tran}. $h_1$ and $h_2$ in Eq. (\ref{SHIT}) stand for the hyperfine mixing coefficient that can be estimated from the ratio of the off-diagonal hyperfine interaction matrix element and the energy difference between the interactive states.

Using the computational model described in Sec. \ref{subsec:CM}, we calculate the hyperfine induced $^3P^o_{0,2} ~-~ ^1S_0$ E1 transition rates and present the results in table \ref{HIT1} and table \ref{HIT2}. Additionally, the off-diagonal hyperfine interaction matrix elements ($W$) and the hyperfine mixing coefficients are displayed as well. It is found from table \ref{HIT1} that the off-diagonal hyperfine interaction matrix elements are well converged with the expansion of the configuration space. While relatively large changes in the hyperfine mixing coefficients between CC, MR and Breit models are mainly attributed to the energy separations involved that are sensitive to the higher-order correlation effects as discussed in Sec.~\ref{subsec:CM}. As can be seen, the final hyperfine induced transition rate is three orders of magnitude larger than the M1 transition presented in Sec. \ref{subsec:M1E2} and thus reduces the lifetime of the states by a factor of 630. Therefore, for $^{21}$Ne the HIT is a dominant decay channel from the $^3P^o_0$ state.

\begin{table}
\caption{\label{HIT1} Hyperfine induced $^3P^o_0 ~-~ ^1S_0$ E1 transition rates $A$ (in s$^{-1}$) for $^{21}$Ne together with off-diagonal hyperfine interaction matrix elements $W$ in (a.u.) and hyperfine mixing coefficients as functions of computational models. The number in square brackets represents the power of 10.}
\begin{tabular}{ccccccccccccccccccccccc}
\hline
\hline
      &  \multicolumn{2}{c}{($^3P^o_1,\; ^3P^o_0$)} && \multicolumn{2}{c}{($^1P^o_1,\; ^3P^o_0$)} & \\
\cline{2-3}\cline{5-6} Model & $W_1$ & $h_1$  && $W_2$ & $h_2$ & $A$ \\
\hline
DF    & $-$1.4241[-7] & $-$8.032[-5]  && $-$1.2290[-7] & 2.668[-5]    &  1.716 \\
3SD   & $-$1.3692[-7] & $-$7.767[-5]  && $-$1.2325[-7] & 2.730[-5]    &  1.644 \\
4SD   & $-$1.1001[-7] & $-$5.383[-5]  && $-$1.3080[-7] & 3.028[-5]    &  1.252 \\
5SD   & $-$1.1566[-7] & $-$5.427[-5]  && $-$1.2899[-7] & 3.082[-5]    &  1.316 \\
6SD   & $-$1.1210[-7] & $-$5.188[-5]  && $-$1.2951[-7] & 3.140[-5]    &  1.323 \\
7fSD  & $-$1.1343[-7] & $-$5.238[-5]  && $-$1.2918[-7] & 3.124[-5]    &  1.328 \\
8fSD  & $-$1.1288[-7] & $-$5.219[-5]  && $-$1.2870[-7] & 3.079[-5]    &  1.312 \\
9fSD  & $-$1.1297[-7] & $-$5.218[-5]  && $-$1.2878[-7] & 3.083[-5]    &  1.316 \\
CC    & $-$1.3475[-7] & $-$6.173[-5]  && $-$1.2337[-7] & 3.000[-5]    &  1.458 \\
MR    & $-$1.3162[-7] & $-$7.072[-5]  && $-$1.2411[-7] & 2.776[-5]    &  1.488 \\
Breit & $-$1.3438[-7] & $-$8.156[-5]  && $-$1.2072[-7] & 2.614[-5]    &  1.484 \\
\hline
\hline
\end{tabular}
\end{table}

For the other hyperfine induced transition from the $^3P^o_2$ state to the ground state, the mechanism is a little more complex since the excited level possesses several hyperfine sublevels with $F=1/2,3/2,5/2,7/2$ for the $^{21}$Ne isotope. Out of them only the $F=1/2,3/2,5/2$ states can decay to the ground state. In table~\ref{HIT2} we present the transition rates and corresponding hyperfine mixing coefficients for these hyperfine states using the Breit model. As can be seen from this table, the HIT rates are somewhat smaller than the M2 transition probability discussed in Sec. \ref{subsec:M2 tran} but still significantly affect the level lifetime.

\begin{table}
\caption{\label{HIT2} F-dependent hyperfine induced $^3P^o_2 ~-~ ^1S_0$ transition rates $A$ (in s$^{-1}$) together with associated hyperfine mixing coefficients h$_1$ and h$_2$ for $^{21}$Ne by using the ``Breit'' model. The number in square brackets represents the power of 10.}
\begin{tabular}{ccccccccccccc}
\hline
\hline
$F$ &   $h_1$     &  $h_2$    &     $A$     \\
\hline
1/2 & $-$4.946[-7] & 5.089[-7] & 2.500[-4] \\
3/2 &    5.779[-6] & 1.351[-6] & 6.395[-5] \\
5/2 &    1.935[-5] & 2.390[-6] & 6.153[-3] \\
\hline
\hline
\end{tabular}
\end{table}

\subsection{\label{subsec:E1 tran} Level lifetimes in $2s^22p^53s$ configuration}
Using the data presented in tables~\ref{E1 tran} - \ref{HIT2} we obtain the lifetimes of states in $2p^53s$ configuration for $^{20,21}$Ne isotopes by
\begin{equation}\label{LT}
\tau_k = \frac{1}{\sum_{i} A_{ki}},
\end{equation}
where the summation is made over the main decay channels. For the $^3P^o_2$ state of $^{21}$Ne isotope, the weighted average lifetime ($\tau = \frac{\sum_i (2F_i + 1)\tau_i}{\sum_i (2F_i + 1)}$) is calculated. The results are reported in table \ref{LL}. It can be seen that the lifetimes of those two metastable states are apparently different owing to the impact of hyperfine interactions, especially for the $^3P^o_0$ state. We should emphasize that the interference effect between the main decay channels is neglected in Eq. (\ref{LT}), which brings about observable variation in lifetimes if transition probabilities have similar orders of magnitude. As discussed in Sec.~\ref{subsec:HIT} the hyperfine induced transition rate of the $^3P^o_2$ state for $^{21}$Ne has the same order of magnitude as the M2 transition, and strong interference may occur. This also influences the radiative emission distribution, which is useful for anisotropy plasma diagnosis~\cite{Yao2006}. Further studies are ongoing.

\begin{table}
\caption{\label{LL} Lifetimes (in s) of levels in $2p^53s$ configuration for $^{20,21}$Ne isotopes. The relevant nuclear parameters are taken from Ref.~\cite{Stone2005}. The number in square brackets represents the power of 10.}
\begin{tabular}{cccccccc}
\hline
\hline
isotope   &  $^3P^o_2$ &  $^3P^o_1$  & $^3P^o_0$  & $^1P^o_1$  \\
\hline
$^{20}$Ne &  17.63     &  1.995[-8]  &  424.1     &  1.638[-9] \\
$^{21}$Ne &  17.10     &  1.995[-8]  &  0.6728    &  1.638[-9] \\
\hline
\hline
\end{tabular}
\end{table}

\subsection{Estimation of uncertainties}
For light atoms such as neon the main uncertainties in calculations of physical quantities arise from electron correlation effects. In this work, large-scale configuration spaces are used to account for these correlation effects in the case of neutral neon, even partly including higher-order correlation among $2s,2p$ valence electrons. The residual higher-order valence correlations and the higher-order correlations between $1s$ core electrons and between core and valence electrons, which are not taken into account, contribute to the uncertainties. By monitoring the convergence of physical quantities under investigation as the active set is enlarged as well as monitoring the changes as the correlation models are defined by including higher-order correlation effects, we estimate that the errors in present calculations is about 2\%. This observation is further strengthened by the excellent agreement between E1 rates in the length and velocity gauges. The hyperfine induced $^3P^o_2 ~-~ ^1S_0$ E1 transition rate is an exception. This transition is sensitive to higher-order correlation effects not included or saturated in our calculations. Moreover, the counteraction between off-diagonal magnetic dipole and electric quadrupole interactions contributes to the uncertainties in this rate. Approximately, these bring about 10\%-20\% error for this transition rate. Other physical effects neglected in this work such as frequency-dependent Breit interactions and quantum electrodynamical (QED) corrections are indeed fractional for neutral neon, as discussed by Avgoustoglou \textit{et al.}~\cite{Avgoustoglou1995}.

\section{Conclusion}
In this work we investigate the transition properties of the main one-photon decay channels for the $2p^53s$ configuration of Ne isotopes using the MCDHF method. The electron correlation effects are taken into account systematically with the active space approach. Detailed comparisons are made with measurements and with other calculations. The effects of Breit interaction on fine structures and transition properties are discussed. It is found that the Breit interaction changes the line strength of the $^3P^o_1 ~-~ ^1S_0$ transition by around 17\%. Present calculations do not resolve the discrepancies in the $^3P^o_2 ~-~ ^1S_0$ M2 transition rates between theories and experiments. Further measurement is therefore called for. The hyperfine induced $^3P^o_{0,2} ~-~ ^1S_0$ E1 transition rates for the $^{21}$Ne isotope are calculated as well. We discovered that the hyperfine interactions drastically affect the lifetime of the metastable states, especially for the $^3P^o_0$ state. The lifetime of states in $2p^53s$ configuration are predicted for both $^{20}$Ne and $^{21}$Ne isotopes.

\section*{Acknowledgement}
This work was supported by the Communaut\'e fran\c{c}aise of Belgium (Action de Recherche Concert\'ee), the Beglian National Fund for Scientific Research (FRFC/IISN Convention) and by the IUAP Belgian State Science Policy (Brix network P7/12). PJ and GG acknowledge support from the Visby program of the Swedish Institute.
\clearpage

\bibliography{Ne}

\end{document}